\documentclass[aps,showpacs, prl, amssymb, amsmath, twocolumn]{revtex4}
\usepackage{amsmath,latexsym,amssymb}
\usepackage{graphicx}
\usepackage{amsmath}
\usepackage{latexsym}
\usepackage{amssymb}
\usepackage{bm}
\usepackage{placeins}
\usepackage{color}
\usepackage{bbm}
\usepackage{multirow}
\usepackage{cancel}
\usepackage{float}
\usepackage{xcolor}
\usepackage{ulem}

\newcommand{\bra}[1]{\left\langle #1\right|}
\newcommand{\ket}[1]{\left|#1\right\rangle}
\newcommand{\abs}[1]{\left|#1\right|}

\begin{document}

\title{Ultrafast time-scale Berry-phase gates of atomic clock states}

\author{Yunheung Song$^{1}$, Jongseok Lim$^{2}$, and Jaewook Ahn$^{1}$}
\email{jwahn@kaist.ac.kr}
\address{$^{1}$Department of Physics, KAIST, Daejeon 305-338, Korea \\ $^{2}$Centre for Cold Matter, Blackett Laboratory, Imperial College London, Prince Consort Road, London SW7 2AZ, United Kingdom}
\date{\today}

\begin{abstract}
Extremely fast qubit controls can greatly reduce the calculation time in quantum computation, and potentially resolve the finite-time decoherence issues in many physical systems. Here, we propose and experimentally demonstrate pico-second time-scale controls of atomic clock state qubits, using Berry-phase gates implemented with a pair of chirped laser pulses. While conventional methods of microwave or Raman transitions do not allow atomic qubit controls within a time faster than the hyperfine free evolution period, our approach of ultrafast Berry-phase gates accomplishes fast clock-state operations. We also achieves operational robustness against laser parametric noises, since geometric phases are determined by adiabatic evolution pathway only, without being affected by any dynamic details. The experimental implementation is conducted with two linearly polarized, chirped ultrafast optical pulses, interacting with five single rubidium atoms in an array of optical tweezer dipole traps, to demonstrate the proposed ultrafast clock-state gates and their operational robustness. 
\end{abstract}

\maketitle

Berry phase is one of the hallmarks in quantum mechanics, dealing with the {\it geometric phase} gained by a quantum wavefunction subjected to an adiabatic process, which can remain nonzero even after a cyclic evolution in which the more familiar dynamic phase disappears~\cite{Berry}. It appears ubiquitously in numerous physical phenomena including Aharonov-Bohm effect, quantum Hall effect, and neutron interferometry, to list a few~\cite{Berry, Klitzing1980, Werner1975}. 

Berry phase written as a unitary operator $U$ for a cyclic evolution is {\it holonomy}, that depends only on the evolution path but not on other dynamic details during the evolution. So, a geometrical manipulation of two-state systems utilizing the Berry phase is expected for robust quantum information processing against environment and parameter noises (characteristically of local nature) due to their independence on local phase changes (dynamic phases), called {\it holonomic quantum gates}~\cite{holonomic,GQCreview}. One way to implement this geometric operation is adiabatic time evolution~\cite{Berry,WZ}. Adiabatic time evolution of a qubit system allows no leakage from an initial adiabatic state of degenerate eigenenergy, so, if an appropriate interaction picture removes this eigenenergy, a parallel transport condition, $\bra{\psi(t)} H_I(t)\ket{\psi(t)}=0$, can be imposed for the holonomy. The time evolution of a qubit system 
driven by the time-varying field of Hamiltonian $ H_I(t)$ 
is written in the bare basis as
\begin{equation}
\ket{\psi(t_f)}=e^{i\phi_{\rm d}} U(\Theta)\ket{\psi(t_i)},
\label{holonomicgate}
\end{equation}
where $\phi_{\rm d}$ is the dynamic phase that is only global, thus ignorable, and $\Theta$ is the geometric phase. 

Original proposals for holonomic quantum gates are based on this adiabatic evolution~\cite{holonomic,atomimp,ionimp,scimp}. However, since it is difficult to satisfy the adiabatic condition for many physical systems of limited coherence time, experimental implementation has been limited to long-lived transitions~\cite{AHQCexp} or the shortcut to adiabaticity~\cite{STAHQCexp,STAHQCexp2}. Most other examples utilize nonadiabatic holonomic quantum gates~\cite{NHQC, NHQCexp,NHQCexp2,NHQCexp3,NHQCexp4}, but nonadiabatic characteristic makes them sensitive to parameter fluctuations~\cite{NHQCerror}.

\begin{figure}[t]
\centerline{\includegraphics[width=0.48\textwidth]{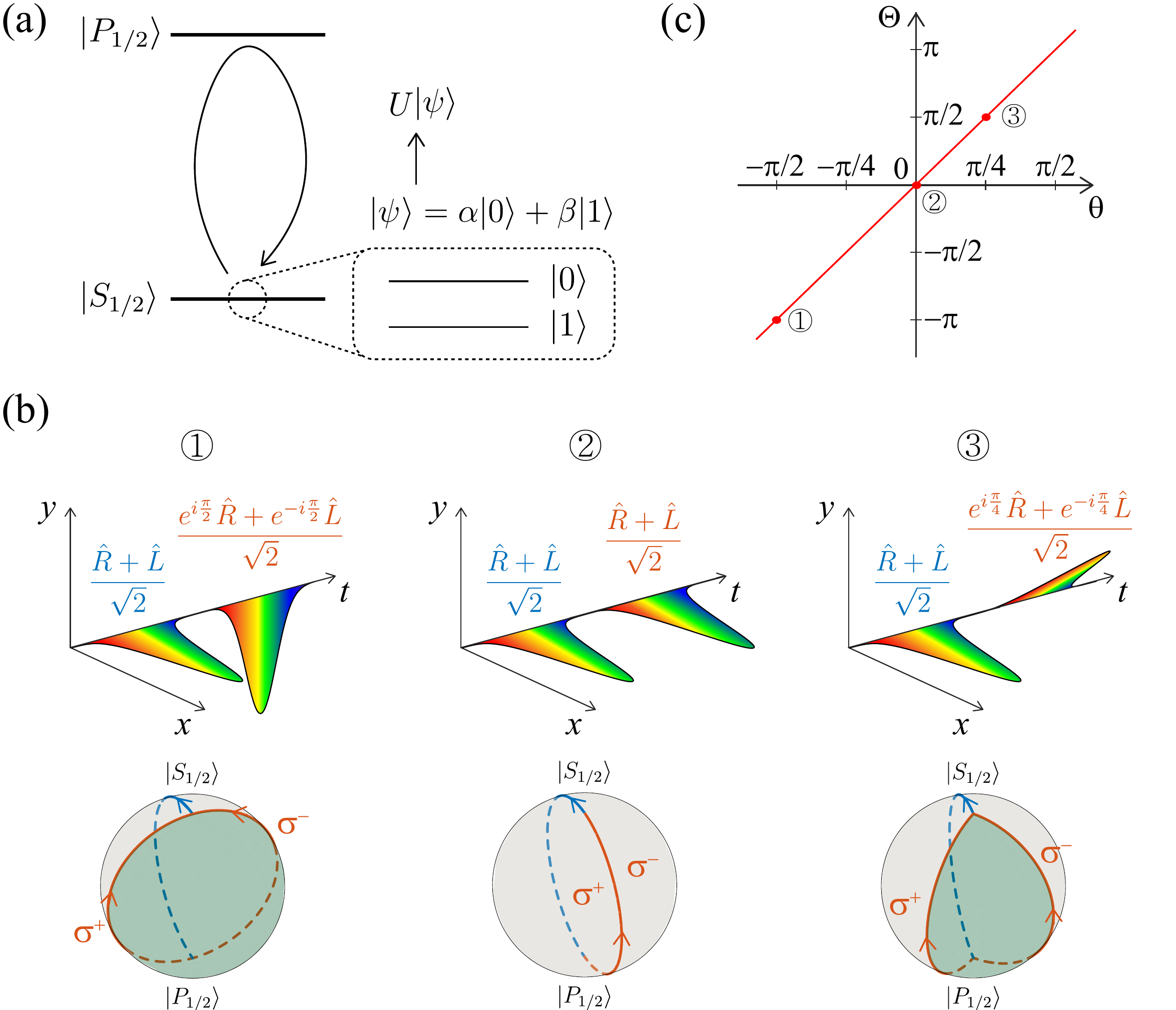}}
\caption{(a) Energy level diagram of a ground level $\ket g$, an excited level $\ket e$, and two degenerate qubit eigenstates $\ket 0$ and $\ket 1$ within the ground level $\ket g$. A cyclic transition between $\ket g$ and $\ket e$ results in the evolution from the initial qubit state $\ket{\psi}$ to $U\ket{\psi}$. (b) The pulse sequence of holonomic gates (upper) and the corresponding time evolutions of the two-level system (lower). When the system undergoes successive adiabatic passages (RAP) by two linearly polarized, chirped laser pulses with relative polarization angle $\theta$ (\raisebox{.5pt}{\textcircled{\raisebox{-.9pt} {1}}} $\theta$=$-\pi$/2, \raisebox{.5pt}{\textcircled{\raisebox{-.9pt} {2}}} 0, \raisebox{.5pt}{\textcircled{\raisebox{-.9pt} {3}}} $\pi$/4),  the first RAP excites the system from $\ket g$ to $\ket e$ along the path shown in blue line, and then the second RAP de-excites the system back to $\ket g$ along the two different paths labeled by $\sigma^{+}$ and $\sigma^{-}$, shown in red lines. The geometric phase gained through the cyclic transitions is proportional to the shaded area enclosed by the two red lines in the Bloch sphere. (c) Geometric phase $\Theta$ vs. relative polarization angle $\theta$, where the three different values of $\theta$ in (b) are indicated with dots.}
\label{fig1}
\end{figure}

In the present paper, we propose a method implementing adiabatic holonomic transitions between atomic clock states within ultrafast time scales. Experimental demonstration is performed on the hyperfine states of an atomic system, interacting with stretched ultrafast optical pulses that allow adiabatic time evolution in the qubit system. We test the robustness of the proposed scheme against laser power fluctuation up to 2,000\% change. We further demonstrate a scheme for rotation operators about arbitrary axes, with which a set of universal one-qubit quantum gates can be constructed.

\begin{figure}[b]
\centerline{\includegraphics[width=0.48\textwidth]{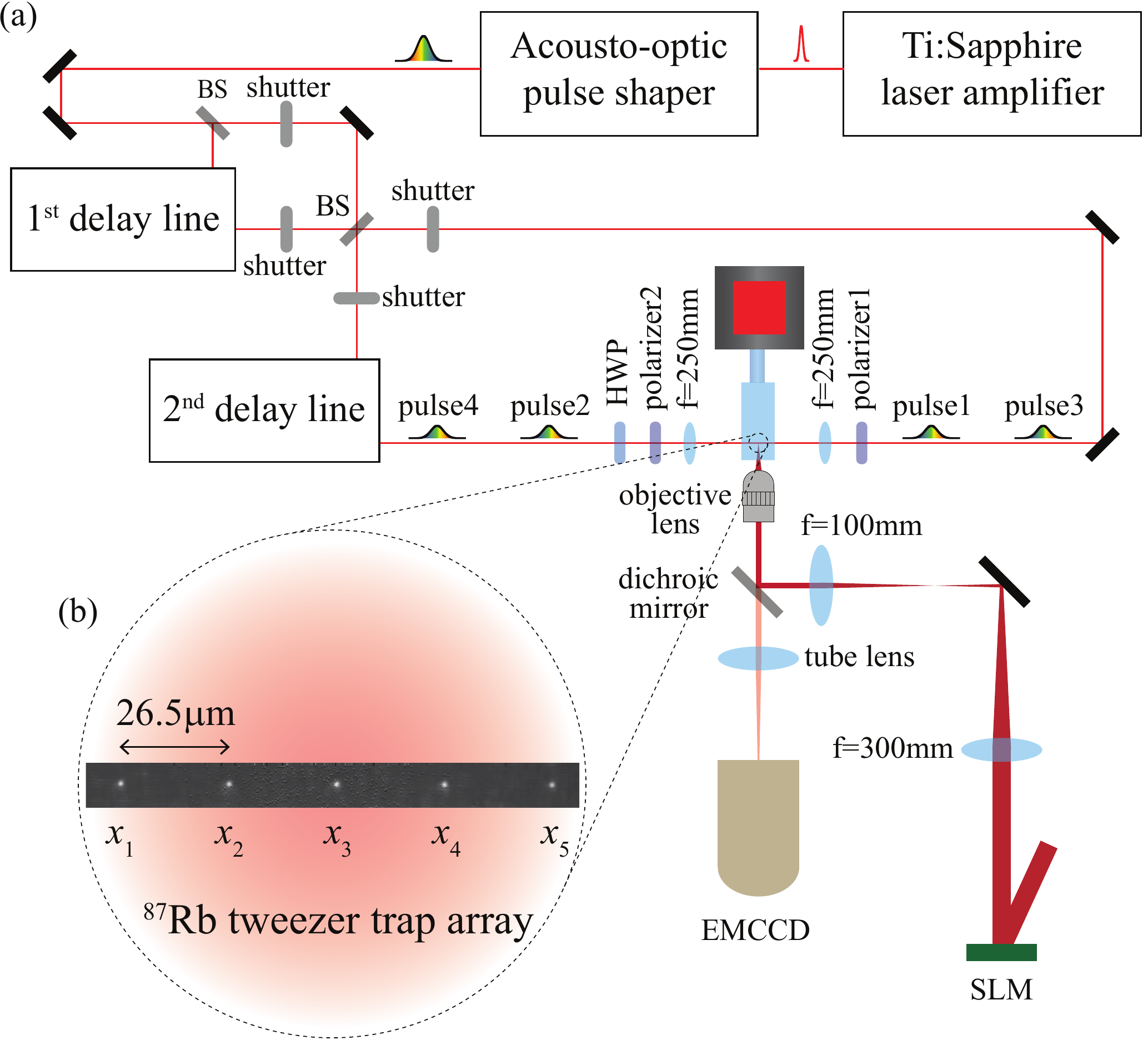}}
\caption{(a) Schematic of the experiment. (b) The fluorescence image of single atoms at positions $x_i$ ($i=1,2,\cdots,5$). (BS: beam splitter, HWP: half-wavelength plate, EMCCD: electron multiplying cathode charge device, SLM: spatial light modulator) }
\label{fig2}
\end{figure}

Let us consider an atomic system (see Fig.~\ref{fig1}a), of an excited level $\ket e=\ket{P_{1/2}}$ and an ground level $\ket g=\ket{S_{1/2}}$, in which $\ket g$ consists of two ground hyperfine states (qubit states) $\ket0=\ket{S_{1/2},F=I+1/2,m_F=0}$ and $\ket1=\ket{S_{1/2},F=I-1/2,m_F=0}$. When the hyperfine energy splitting $\hbar\omega_{\rm hf}$ is negligible compared to the inverse of gate-operation time, these qubit states can be considered as energy degenerate states. We utilize chirped ultrafast optical pulses to implement a rapid adiabatic passage (RAP)~\cite{RAPreview} that provides robust adiabatic population transfer between $\ket g$ and $\ket e$. Successive RAP applications by a pair of chirped pulses (see Fig.~\ref{fig1}b) adiabatically drives the transition from the ground initial state $\ket{\psi(t_i)}=\alpha\ket0+\beta\ket1$ to $\ket e$ by the first pulse and then back to the ground state by the second, making a cyclic time evolution. Suppose the pulses propagate along the quantization axis ($+\hat{z}$ axis) of the qubit states $\ket0$ and $\ket1$, and are linearly polarized with a relative polarization angle $\theta$ between them. We set the coordinate system to let the polarization unit vector of the first pulse be the $\hat{x}$-axis unit vector $\hat x=(\hat R+\hat L)/{\sqrt2}$, where $\hat R$ and $\hat L$ are the right and left circular polarization unit vectors, respectively. Then, the polarization unit vector of the second pulse is expressed as $(e^{-i\theta}\hat R+e^{i\theta}\hat L)/\sqrt{2}$. Correspondingly, the qubit system is considered in the `Cartesian' basis, as $\ket{\psi(t_i)}=\frac{\alpha-\beta}{\sqrt{2}}\ket-+\frac{\alpha+\beta}{\sqrt{2}}\ket+$, where $\ket\pm\equiv(\ket0\pm\ket1)/{\sqrt2}$ are the fine-structure states $\ket{S_{1/2},m_J=\pm1/2}$ in our case. By the dipole selection rule, the right and left circular polarizations drive $\sigma^{\pm}$ transitions between $\ket\mp$ and $\ket e$, respectively. In the lower figures of Fig.~\ref{fig1}(b), Bloch sphere representation shows the time evolution pathways of $\ket\mp$ driven by respective polarization components. After the cyclic evolution by the two pulses, $\ket\mp$ states get geometric phase $\pm\theta-\pi$, respectively, corresponding to minus half of the solid angle enclosed by the evolution pathways~\cite{GQCreview}, while dynamic phase $\phi_{\rm d}$ is due to intensity- and detuning-dependent eigenenergy~\cite{RAPreview} and dynamic Stark shift from neighboring transitions~\cite{ultrafastXrot}. So we get  $\ket{\psi(t_f)}=\frac{\alpha-\beta}{\sqrt{2}} e^{i\phi_{-}} \ket- + \frac{\alpha+\beta}{\sqrt{2}} e^{i\phi_{+}} \ket+$, in which $\phi_{\mp}=\pm\theta-\pi+\phi_{\rm d}$ are the phases gained along the time evolution. The final state $\ket{\psi(t_f)}$ is then expressed in the qubit basis as
\begin{eqnarray}
\ket{\psi(t_f)}&=&-e^{i\phi_{\rm d}}
\begin{pmatrix}
\cos\theta & -i\sin\theta \\
-i\sin\theta & \cos\theta
\end{pmatrix}
\ket{\psi(t_i)}\nonumber\\
&=& -e^{i\phi_{\rm d}} U_{\hat x}(\Theta)\ket{\psi(t_i)},
\label{ultrafastholonomic}
\end{eqnarray}
where $\Theta=2\theta$ and $U_{\hat x}(\Theta)$ is the $X$-rotation operator of the qubit states by angle $\Theta$  (see Fig.~\ref{fig1}c). Note that the dynamic phase $\phi_{\rm d}$ is always global because linear polarization guarantees an equal magnitude of the $\sigma^{\pm}$ transitions and thus the same dynamic phase for each transition. Therefore, this scheme implements the holonomic transition determined by only geometric phase $\Theta$, robust against laser parameters such as intensity and detuning.

\begin{figure*}[t]
\centerline{\includegraphics[width=0.9\textwidth]{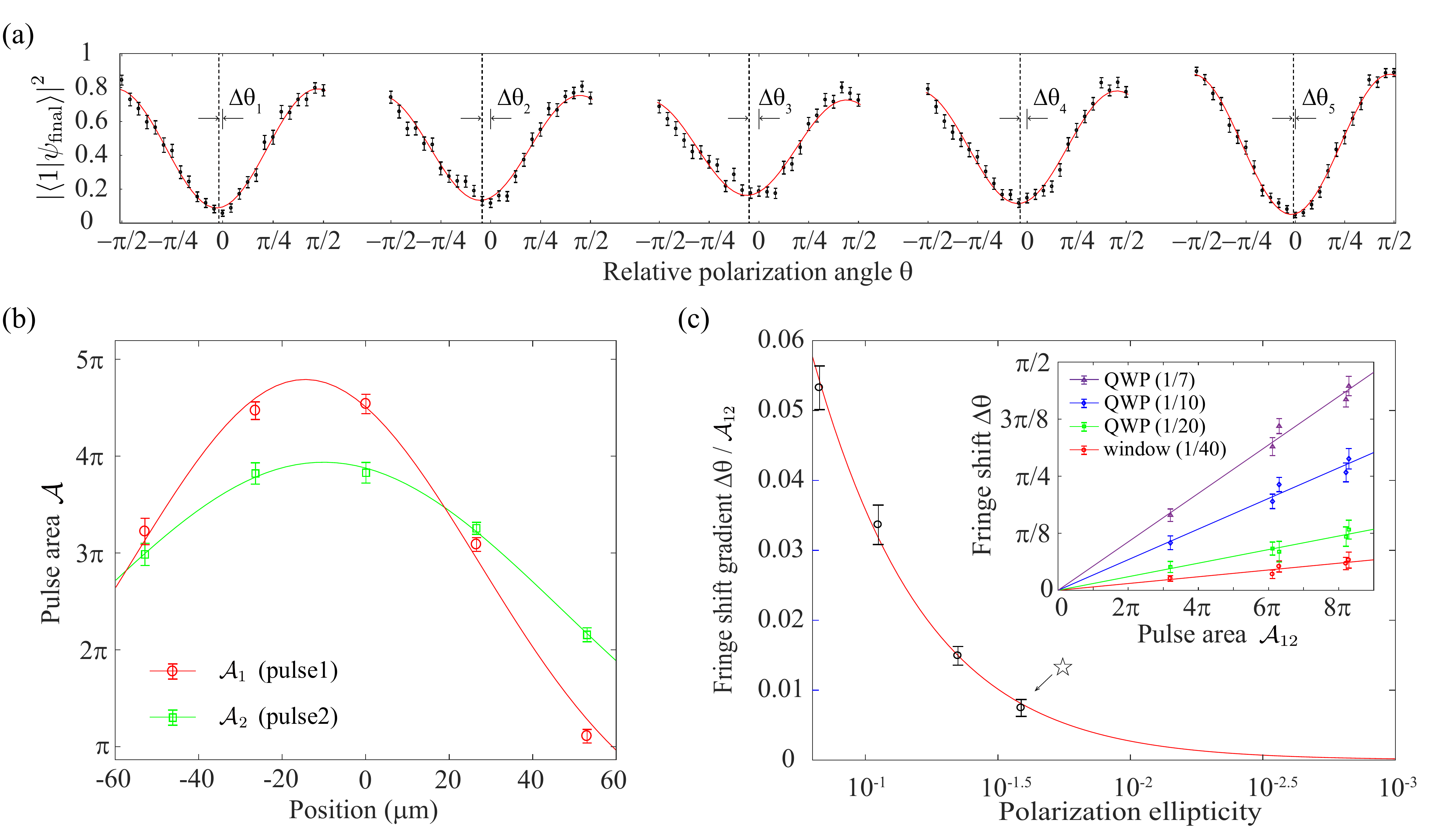}}
\caption{(a) The $\ket 1$ state probability of the qubit, prepared in the $\ket 0$ state followed by the $U_{\hat x}$ operation. Each plot shows the probability measured from each single atom at positions $x_{i}$ for various relative polarization angle $\theta$. Black dots are the measurements and the red lines are their fits to a sinusoidal function (see supplements for details). (b) Pulse area at each trap site for pulse1 (red circles) and pulse2 (green squares), and their fits to a Gaussian function (red and green lines, respectively). (c) The gradient of fringe shift vs. polarization ellipticity, for four different values of polarization ellipticity of the ultrafast pulses: 1/40 (induced by the vacuum window; 1/20, 1/10, and 1/7 (varied by a quarter-wave plate). The polarization ellipticity induced in the experiment is indicated by the star mark. Inset: The fringe shifts $\Delta\theta$ as a function of the pulse area for the four values of polarization ellipticity.}
\label{fig3}
\end{figure*}

Qubit rotations about an arbitrary axis, $\hat n=n_x\hat{x} + n_y\hat{y}$, can be implemented with an additional pair of time-delayed pulses.
Since our scheme works in the regime where the hyperfine splitting is neglected, we adopt a method utilizing the hyperfine interaction in a longer time scale~\cite{ultrafastXrot}. In the interaction picture where the qubit basis is $\ket{0'}=\ket0$ and $\ket{1'}=e^{-i \omega_{\rm hf}t}\ket1$, the `Cartesian' basis is given by $\ket\pm=\ket{0'}\pm e^{i \omega_{\rm hf}t}\ket{1'}$. Then, with the second Berry-phase gate applied after time delay $T$, the time evolution of the qubit system from $t_i+T$ to $t_f+T$ becomes $\ket{\psi(t_f+T)}=-e^{i\phi_{\rm d}}U_{\hat n}(\Theta)\ket{\psi(t_i+T)}$, where
\begin{equation} U_{\hat n}(\Theta)=
\begin{pmatrix}
\cos\theta & -i(n_x+in_y)\sin\theta \\
-i(n_x-in_y)\sin\theta & \cos\theta
\end{pmatrix}
\end{equation}
is the rotation operator of the qubit states about the axis $\hat n$, with $n_x=\cos(\omega_{\rm hf}T)$ and $n_y=\sin(\omega_{\rm hf}T)$ controlled by the time delay $T$.

Experimental demonstration of the ultrafast {Berry-phase} gates was performed with an array of single rubidium-87 atoms ($^{87}$Rb) driven by pulse-shaped sub-picosecond laser pulses (see Fig.~\ref{fig2}a). Femtosecond laser pulses were produced by a femtosecond Ti:sapphire amplifier system operated at 1-kHz repetition rate (carrier frequency $377.1$~THz, bandwidth $3.8$~THz), which were resonant to the D$_1$ transition. The pulses were linearly chirped by an acousto-optic pulse shaper to stretch the pulse length to 1.5~ps with a chirp rate of 2.6~ps$^{-2}$, to satisfy the adiabatic condition for the RAP. Each pulse was split into two pairs of double pulses, with the inter-pair (intra-pair) delay {$T=70-370$~ps ($\tau=6.7$~ps)}. The relative polarization {angle} $\theta$ was varied by a combination of a half-wave plate and a polarizer, realizing $ U_{\hat x}$ with the first pair and $U_{\hat n}$ with the second. These pulses were delivered along the counter-propagating directions ($\pm \hat{z}$) to the array of single atoms trapped in a vacuum chamber (see Fig.~\ref{fig2}b). Five single $^{87}$Rb atoms were prepared at fixed positions with 26.5~$\mu$m spacing, along the direction perpendicular to the laser beam propagation axis, by optical tweezers in a magneto-optical trap~\cite{Ashkin1970,DHOT}. The optical tweezers were tightly focused 852~nm laser beams (2~$\mu$m $1/e^2$ diameter) with the trap depth of 1.6~mK. The atoms were first optically pumped to the $\ket0\equiv\ket{5S_{1/2},F=2,m_F=0}$ qubit state using $\pi$-polarized continuous light resonant with the $F=2\rightarrow F'=2$ transition of the D$_1$ line and the $F=1\rightarrow F'=2$ transition of the D$_2$ line, in the presence of applied magnetic field of 2.4~G which defined the quantization axis along the laser propagation axis. Then, the ultrafast pulse sequence, each pair of which constituted one {Berry-phase gate} operation, was focused to the single-atom array with the beam waist of 60~$\mu$m (90~$\mu$m) for {pulse1 and pulse3 (pulse2 and pulse4)} which was smaller than the array size of 106~$\mu$m. Thus each atom in the array experienced largely different intensities. Finally, push-out measurement~\cite{pushout} was applied to record the probability of the $\ket1=\ket{5S_{1/2},F=1,m_F=0}$ state of each atom with an EMCCD camera.

With the experimental apparatus, we first demonstrate the robustness of $U_{\hat x}(\Theta)$ in Eq.~\eqref{ultrafastholonomic} against laser power fluctuation. We used the first pair of the pulses (pulse1 and pulse2 in Fig.~\ref{fig2}a), while blocking the second pair, and measured the state $\ket1$ probability, $|\langle{1}\ket{\psi(t_f)}|^2$, of each atom as a function of the relative polarization {angle} $\theta$. The experimental results are plotted in Fig.~\ref{fig3}a, in which, although the atoms were exposed to different, position-dependent pulse-areas ($\mathcal{A}_{\rm max}\approx 5 \times \mathcal{A}_{\rm min}$, see Fig.~3b), the measured $\Theta$s exhibit very little error {$\Delta\Theta<0.4$~(rad)}. This error mainly came from the birefringence of the vacuum window~\cite{vacwindow}, which affected the polarization of the laser pulses and the imbalance between the $\sigma^{\pm}$ transitions causing dynamic phase error. This polarization imperfection was verified by measuring the gradient of the fringe shift, {$\Delta\Theta/\mathcal{A}$}, vs. polarization ellipticity, as shown in Fig.~\ref{fig3}c. In our current demonstration of the $U_{\hat x}$ which was limited by remaining polarization ellipticity of 1/40, the robustness against the laser intensity is achieved up to {$\Delta \Theta / \mathcal{A} = 1.5$\%}, and the ultra-low birefringence technique~\cite{ultralowbirefringence} of 1/3000 ellipticity is expected to further improve this below $\sim 0.01$\%.

\begin{figure}[thb]
\centerline{\includegraphics[width=0.45\textwidth]{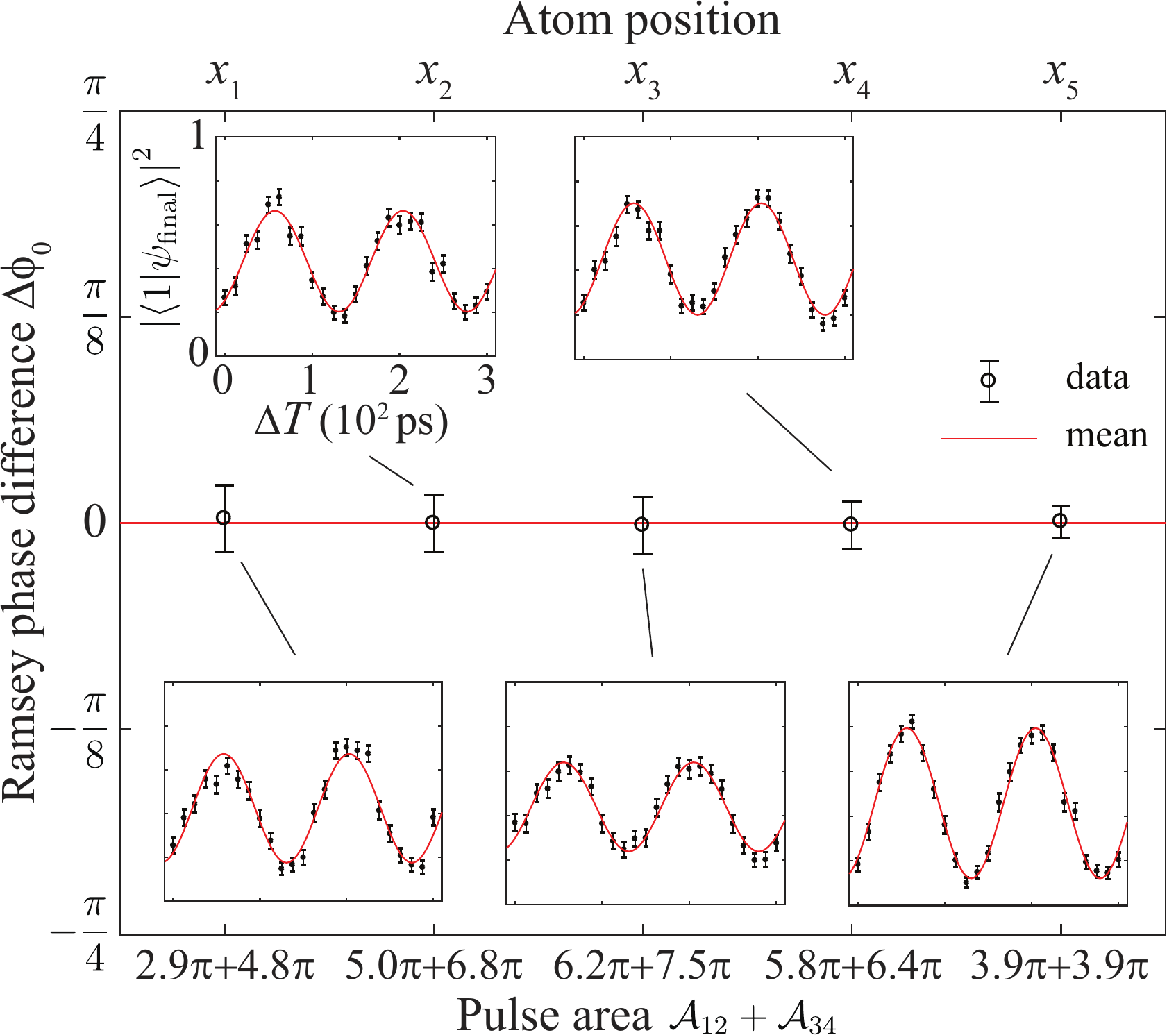}}
\caption{The Ramsey phase difference $\Delta\phi_{0}$, which is defined as the deviation of the Ramsey phase at each atom from the mean value of it. As in the inset graphs, the Ramsey fringe at each atom position (and each corresponding total pulse area) is measured by the transition probability from $\ket0$ to $\ket1$ after applying ${U}_{\hat{x}}(\pi/2)$ (pulse1 and pulse2) and ${U}_{\hat{n}}(\pi/2)$ (pulse3 and pulse4) with various inter-pair time delay $T$, consecutively. Fitted values of $\Delta\phi_{0}$ are plotted with the errorbars indicating the 95\% confidence intervals.}
\label{fig4}
\end{figure}

In the second experiment, we test the robustness of the rotational axis $\hat{n}$, using two {Berry-phase gates $U_{\hat x}$ (pulses 1 and 2) and $U_{\hat n}$ (pulses 3 and 4).} The relative polarization {angle} $\theta$ of both gates was fixed to $\pi/4$ for maximum visibility, and the Ramsey fringe of the $F=1$ state probability was measured, with respect to the time delay $T$ between the pulse pairs, as 
\begin{equation}
\abs{\bra1 U_{\hat n}(\Theta)~U_{\hat x}(\Theta)\ket0}^2=\sin^2\Theta \cos^2(\phi_{\hat{n}}+\phi_0),
\end{equation} 
where $\phi_{\hat{n}}=\omega_{\rm hf}T/2$ is the angle of the rotational axis and $\phi_0$ is a constant. Ramsey phase differences, $\Delta\phi_{0}=\phi_{0}-\left<\phi_{0}\right>$, are shown in Fig.~\ref{fig4}, with
the Ramsey fringe of each atom in the inset. The measured frequency of the Ramsey fringe at each atom also agrees with the $^{87}$Rb ground hyperfine splitting $\omega_{\rm hf}$ within the 95$\%$ confidence interval, and the mean value is $2\pi \times 6.79\pm0.08$~GHz. In order to estimate how robust the rotation axis of the Berry-phase gate is, we consider the possibility of intensity-dependent axis shift $\phi_0(\mathcal{A})$, i.e., $\hat n(T)\rightarrow \hat n'(T,\mathcal{A})$ with $n'_x(T,\mathcal{A})=\cos(\phi_{\hat{n}}+\phi_0(\mathcal{A}))$ and $n'_y(T,\mathcal{A})=\sin(\phi_{\hat{n}}+\phi_0(\mathcal{A}))$. However, the Ramsey fringes in the insets of Fig.~\ref{fig4} show sinusoidal curves of the same phase shift, for all the atom positions regardless of the pulse areas, demonstrating the robustness in $\hat{n}$. In other words, all the values of $\Delta\phi_{0}$ are zero within 95$\%$ confidence intervals among all the atom positions. The mean value of the confidence interval radius for all the atom positions is $0.016\pi$ while the standard deviation of the pulse area among all the gates is $1.45\pi$, showing the intensity robustness within their ratio of $1.1\%$.

In conclusion, we have implemented qubit rotations of atomic clock states using Berry phases induced by two linearly-polarized chirped pulses in picosecond time scales. We note that conventional implementations of holonomic gates using spectroscopic distinguishability cannot be applied in these time scales. The results show, as a characteristic of geometric phases of adiabatic passages, gate operation robustness against laser parameter errors, which has been hard to achieve in previous nonadiabatic holonomic schemes. The ultrafast Berry-phase gates can offer a fast and robust qubit control not only for atomic systems, but also for solid state systems of relatively short coherence time.

\begin{acknowledgements}
This research was supported by Samsung Science and Technology Foundation [SSTF-BA1301-12].
\end{acknowledgements}

\section{Supplemental Material}

\subsection{Detailed description of the ultrafast Berry-phase gates}

The proposed Berry-phase gates for the atomic clock states are achieved with two successive chirped ultrafast pulses that are linearly polarized or in equal magnitudes of left and right circular polarizations. Each circular polarization component of the chirped pulses adiabatically drives population transfer between the fine-structure ground and excited levels according to the transition selection rules. The Berry-phase difference between the two driven evolution paths is determined by the relative polarization angle between the two pulses, resulting in robust qubit rotation insensitive to other laser parameters except the polarization. We first briefly review the chirped rapid adiabatic passage and then describe the Berry-phase gates for atomic systems.

\subsubsection{Chirped rapid adiabatic passages} \label{CRAP}
The robust population transfer is implemented by using the chirped rapid adiabatic passage (RAP)~\cite{RAPreview}. Let us consider a two-level system, of the fine-structure states $\ket{g}$ and $\ket{e}$ with energy separation $\hbar\omega_0$, is interacted with a Gaussian chirped pulse of an electric field written in the frequency domain as\

\begin{eqnarray}
E(\omega) =   \frac{E_p}{2}e^{-{(\omega-\omega_p)^2}/{\Delta \omega_p^2}} \times e^{-i{c_p}(\omega-\omega_p)^2/2}+c.c.,
\end{eqnarray}
where $E_p$ is the peak amplitude, $\omega_p$ is the laser frequency, $\Delta\omega_p$ is the bandwidth, and $c_p$ is the chirp parameter~\cite{WeinerBook}. The corresponding time-domain electric field is given by

\begin{eqnarray}
E(t) =  \frac{\mathcal{E}_p}{2} e^{-{t^2}/{\Delta t_p^2}} \times e^{-i(\Gamma_p t ^2+\omega_p t + \varphi_p)} +  c.c.
\label{Efield}
\end{eqnarray}
with $\mathcal{E}_p = E_p\sqrt{{\Delta\omega_p}/{\Delta t_p}} $, $\Delta t_p=\sqrt{4/\Delta\omega_p^2+c_p^2\Delta\omega_p^2}$, $\Gamma_p = {c_p}/(2c_p^2+8/\Delta\omega_p^4)$, and $\varphi_p =-\tan^{-1}({c_p\Delta\omega_p^2}/{2})/2$. The Hamiltonian of this interaction is given by
\begin{eqnarray}
{H}=\frac{\hbar}{2}\begin{pmatrix}
-\Delta(t) & \Omega(t) \\
\Omega(t) & \Delta(t)
\end{pmatrix}
\label{Hamiltonian}
\end{eqnarray}
in the interaction picture basis \{$\ket{g'}=e^{-i\int_{-\infty}^t dt'\Delta(t')/2}\ket g$, $\ket{e'}=e^{-i\left(-\int_{-\infty}^t dt'\Delta(t')/2+\omega_0 t+\varphi_p\right)}\ket e$\}, where $\Delta(t)=\omega_0-\omega_p-2\Gamma_p t$ is the detuning, $\Omega(t)=-\mu\mathcal{E}_p e^{-{t^2}/{\Delta t_p^2}}/\hbar$ is the Rabi frequency, and $\mu$ is the transition dipole moment. The eigenstates of Eq.~\eqref{Hamiltonian} are given by
\begin{eqnarray}
|\epsilon_+(t)\rangle &=& \sin\vartheta(t)\ket{g'} +\cos\vartheta(t)\ket{e'} \\
|\epsilon_-(t)\rangle &=& \cos\vartheta(t)\ket{g'} -\sin\vartheta(t)\ket{e'} 
\end{eqnarray}
with $\vartheta(t) = \tan^{-1} ({\Omega(t)}/{\Delta(t)})/2$ (for $0 \le\vartheta(t)\le{\pi}/{2}$), and the corresponding eigenenergies are
\begin{eqnarray}
\epsilon_\pm(t) = \pm \frac{\hbar}{2}\sqrt{\Omega^2(t)+\Delta^2(t)}.
\end{eqnarray}
 Here, since the detuning $\Delta(t)$ is linearly dependent on time,  the eigenstate $\ket{\epsilon_-(t)}$ ($\ket{\epsilon_+(t)}$) evolves from $\ket g$ ($\ket e$) to $\ket e$ ($\ket g$) as time changes from $t=-\infty$ to $\infty$, along the meridian of the Bloch sphere. Thus, the complete population transfer between $\ket{g}$ and $\ket{e}$ is achieved as
\begin{eqnarray}
\ket{g}&\rightarrow& -e^{i\left(\frac{1}{2}\int_{-\infty}^{\infty}dt\left(\Delta+\sqrt{\Omega^2+\Delta^2}\right)-\omega_0 t-\varphi_p\right)}\ket{e} \label{adia1} \\
\ket{e}&\rightarrow& e^{i\left(-\frac{1}{2}\int_{-\infty}^{\infty}dt\left(\Delta+\sqrt{\Omega^2+\Delta^2}\right)+\varphi_p\right)}\ket{g}\label{adia2}
\end{eqnarray}
when the adiabatic condition
\begin{equation}
\frac{\left|\dot{\Omega}(t)\Delta(t)-\Omega(t)\dot{\Delta}(t)\right|}{2({\Omega^2+\Delta^2})^{3/2}}=\frac{\Gamma_p|\Omega(t)|\left(2t^2/\Delta t_p^2+1\right)}{
(|\Omega(t)|^2 + 4\Gamma_p^2t^2)^{3/2}} \ll1
\label{adiacondition}
\end{equation}
is satisfied. The rapid adiabatic passage ensures the robustness against the fluctuation of the laser parameters $E_p$, $\omega_p$, and $\varphi_p$ (amplitude, frequency, and phase).

\subsubsection{Description of the ultrafast Berry-phase gates in atomic systems}

In our consideration, the qubit states are the hyperfine states, $\ket{0}=\ket{S_{1/2}, F=I+1/2, m_F}$ and $\ket{1}=\ket{S_{1/2}, F=I-1/2, m_F}$ of the ground state $\ket{g}=\ket{S_{1/2}, m_J=\pm1/2}$, and the excited level is the $\ket{e}=\ket{P_{1/2}, m_J=\pm1/2}$ of an alkali atom. Berry phase gates are implemented by successive optical transitions between $\ket{g}$ and $\ket{e}$, which induce the phase gates for the qubit system of $\ket{0}$ and $\ket{1}$ (atomic clock states for $m_F=0$).

Let us consider two chirped pulses 1 and 2 that are time-separated by $\tau$ and propagating along the $\hat{z}$ axis, of which the total electric field is given by

\begin{widetext}
\begin{eqnarray}
\vec{E}(t) &=&  \hat{n}_1 E_1 \left(t-{\tau}/{2}\right) +  \hat{n}_2 E_2 \left(t+{\tau}/{2}\right)+c.c. \nonumber \\
&=& \left(E^+_1(t)\hat R+E^-_1(t)\hat L\right)+\left(E^+_2(t)\hat R+E^-_2(t)\hat L\right) + c.c.,
\end{eqnarray}
where $\hat n_j= \hat{x} \cos{\theta_j}  +  \hat{y} \sin{\theta_j}$ ($j=1,2$) are the polarization vectors of the pulse $j$, and $E^\pm_j(t)=e^{\mp i\theta_j}E_j\left(t-(-1)^j \tau/2\right)$ are the corresponding electric field components for circular polarizations, $\hat R=(\hat x+i\hat y)/\sqrt2$ and $\hat L=(\hat x-i\hat y)/\sqrt2$, respectively. The interaction Hamiltonian, $H_{\rm int}=-\vec\mu\cdot\vec E$, has no dependence on $I$ in the ultrafast time scale, so the coupling for each polarization component of each pulse is given as an independent two-level system, i.e.,

\begin{align}[S1]
&\bra{P_{\frac{1}{2}},m'_J, I,m'_I} H_{\rm int}\ket{S_{\frac{1}{2}},m_J, I,m_I} 
=\bra{P_{\frac{1}{2}},m'_J} H_{\rm int}\ket{S_{\frac{1}{2}},m_J}\left<{I,m'_I}|{I,m_I}\right>\nonumber\\
&~~~=\sum_{j=1,2}\left(\bra{P_{\frac{1}{2}},\frac{1}{2}}-\vec\mu\cdot\hat R E^+_j(t)\ket{S_{\frac{1}{2}},-\frac{1}{2}}\delta_{m_J',m_J+1}+\bra{P_{\frac{1}{2}},-\frac{1}{2}}-\vec\mu\cdot\hat L E^-_j(t)\ket{S_{\frac{1}{2}},\frac{1}{2}}\delta_{m_J',m_J-1}\right)\delta_{m'_I,m_I}.
\end{align}
In the fine-structure basis, $ \ket{S_{1/2}, F=I\pm1/2,m_F}
=\sum_{\substack{m_J=\pm1/2}}C^{{1}/{2},I,I\pm1/2}_{m_J,m_F-m_J}\ket{S_{1/2},m_J}\ket{I,m_I=m_F-m_J}$,
the time evolution of the ground-hyperfine-state pair $\ket{S_{1/2}, F=I\pm1/2,m_F}$ can be described by the time evolution of the two sets of two-level systems, $\{\ket{S_{1/2},m_J=-1/2}\ket{I,m_F-1/2},\ket{P_{1/2},m_J=1/2}\ket{I,m_F-1/2}\}$ and $\{\ket{S_{1/2},1/2}\ket{I,m_F+1/2},\ket{P_{1/2},-1/2}\ket{I,m_F+1/2}\}$, for  $m_F=0,\pm1$. Note here that $m_F=\pm2$ states do not form a pair of hyperfine states, so we will consider only $m_F=0,\pm1$. 

For $t\le0$ (the first pulse case, $j=1$), the time evolution from the initial ground hyperfine state $\ket{g}=\ket{S_{1/2},m_J=\mp1/2}$ to the excited state $\ket{e}=\ket{P_{1/2},m_J=\pm1/2}$ is a rapid adiabatic passage, as described by Eq.~\eqref{adia1}, when the time separation is long enough to satisfy the adiabatic condition, i.e., $\tau\gg1/\Gamma_p$. So the initial state $\ket{S_{1/2}, F=I\pm1/2,m_F}$ evolves to
\begin{eqnarray}
-C^{\frac{1}{2},I,I\pm1/2}_{-1/2,m_F+1/2}\exp\left({i\left(\frac{1}{2}\int_{-\infty}^0 dt\left(\Delta_1(t)+\sqrt{\Omega^{+}_{1}(t)^2+\Delta_1(t)^2}\right)-\omega_0t-\varphi_1-\theta_1\right)}\right)\ket{P_{1/2},1/2}\ket{I,m_F+1/2} &&  \\
-C^{\frac{1}{2},I,I\pm1/2}_{1/2,m_F-1/2}\exp\left({i\left(\frac{1}{2}\int_{-\infty}^0 dt\left(\Delta_1(t)+\sqrt{\Omega^{-}_{1}(t)^2+\Delta_1(t)^2}\right)-\omega_0t-\varphi_1+\theta_1\right)}\right)\ket{P_{1/2},-1/2}\ket{I,m_F-1/2} &&
\end{eqnarray}
where $\Omega^{+}_{1}(t)= \bra{P_{1/2},1/2} -\vec{\mu} \cdot \hat{R} \ket{S_{1/2},-1/2} \abs{E^+_1(t)} /{\hbar}$, $\Omega^{-}_{1}(t)= \bra{P_{1/2},-1/2} -\vec{\mu} \cdot \hat{L} \ket{S_{1/2},1/2}\abs{E^-_1(t)} /{\hbar}$, and $\Delta_1(t)=\omega_0-\omega_p-2\Gamma_p\left(t+\tau/2\right)$. 

For $t\ge0$ (the second pulse case, $j=2$), the subsequent adiabatic passage from $\ket{e}$ back to $\ket{g}$, according to Eq.~\eqref{adia2}, results in
\begin{align}
&C^{\frac{1}{2},I,I\pm1/2}_{-1/2,m_F+1/2}e^{i\phi_-}\ket{S_{1/2},-1/2}\ket{I,m_F+1/2}+C^{\frac{1}{2},I,I\pm1/2}_{1/2,m_F-1/2}e^{i\phi_+}\ket{S_{1/2},1/2}\ket{I,m_F-1/2}\nonumber\\
&~=e^{i(\phi_-+\phi_+)/2}\sum_{\substack{k=\pm1/2}}C^{\frac{1}{2},I,I\pm1/2}_{-k,m_F+k}e^{ki(\phi_--\phi_+)}\ket{S_{1/2},-k}\ket{I,m_F+k}\nonumber\\
&~=e^{i(\phi_-+\phi_+)/2}\sum_{\substack{k,l=\pm1/2}}C^{\frac{1}{2},I,I\pm1/2}_{-k,m_F+k}C^{\frac{1}{2},I,I+l}_{-k,m_F+k}e^{ki(\phi_--\phi_+)}\ket{S_{1/2}, F=I+l,m_F}\nonumber\\
&~=e^{i(\phi_-+\phi_+)/2} ~ U(\phi_--\phi_+) \ket{S_{1/2}, F=I\pm1/2,m_F}, \label{eq12}
\end{align}
where the total phases, for the $\pm$ polarization components, gained during the two adiabatic evolutions are, respectively,
\begin{eqnarray}
\phi_-&=&\frac{1}{2}\int_{-\infty}^0 dt\left(\Delta_1+\sqrt{{\Omega^{+}_{1}}^2+\Delta_1^2}\right)-\frac{1}{2}\int^{\infty}_0 dt\left(\Delta_2+\sqrt{{\Omega^{+}_{2}}^2+\Delta_2^2}\right)-\varphi_1+\varphi_2-\theta_1+\theta_2 \label{eq13} \\
\phi_+&=&\frac{1}{2}\int_{-\infty}^0 dt \left(\Delta_1+\sqrt{{\Omega^{-}_{1}}^2+\Delta_1^2}\right)-\frac{1}{2}\int^{\infty}_0 dt\left(\Delta_2+\sqrt{{\Omega^{-}_{2}}^2+\Delta_2^2}\right)-\varphi_1+\varphi_2+\theta_1-\theta_2 \label{eq14} 
\end{eqnarray}
with $\Omega^{+}_{2}(t)=\bra{P_{1/2},1/2} -\vec{\mu} \cdot \hat{R} \ket{S_{1/2},-1/2} \abs{E^+_2(t)}/{\hbar}$, $\Omega^{-}_{2}(t)= \bra{P_{1/2},-1/2} -\vec{\mu} \cdot \hat{L} \ket{S_{1/2},1/2}\abs{E^-_2(t)} /{\hbar}$, and $\Delta_2(t)=\omega_0-\omega_p-2\Gamma_p\left(t-\tau/2\right)$. The resulting unitary operations in Eq.~\eqref{eq12} are rotations, respectively, given by
\begin{eqnarray}
U(\Theta)&=&\begin{pmatrix} \cos\frac{\Theta}{2} & -i\sin\frac{\Theta}{2} \\ -i\sin\frac{\Theta}{2} & \cos\frac{\Theta}{2} \end{pmatrix} \quad{\rm and}\quad
\begin{pmatrix} \cos\frac{\Theta}{2}-i\frac{1}{2}\sin\frac{\Theta}{2} & \mp i\frac{\sqrt3}{2}\sin\frac{\Theta}{2} \\ \mp i\frac{\sqrt3}{2}\sin\frac{\Theta}{2} & \cos\frac{\Theta}{2}+i\frac{1}{2}\sin\frac{\Theta}{2} \end{pmatrix}, \quad{\rm for}~m_F=0,\pm1,
\label{rot}
\end{eqnarray}
which correspond to $U_{\hat x}(\Theta)$ and $ U_{\cos({\pi}/{3})\hat z\pm\sin(\pi/3)\hat x}(\Theta)$. For reference, the rotations for $m_F=\pm2$ are the identity. 
\end{widetext}

Therefore, the two chirped pulses rotate the ground two-level system, $\ket{0_{m_F}}\equiv\ket{S_{1/2},F=I+1/2,m_F}$ and $\ket{1_{m_F}}\equiv\ket{S_{1/2},F=I-1/2,m_F}$, by inducing the relative phase between them. Note that since $\Omega^+_j(t)^2=\Omega^-_j(t)^2$ is satisfied due to the symmetry between $m_J=\pm1/2$ and the linear-polarization condition ($\abs{E^+_j(t)}=\abs{E^-_j(t)}$), the dynamic phases represented by the integrals in Eqs.~\eqref{eq13} and \eqref{eq14} in $\phi_+$ and $\phi_-$ are the same. The relative phase is therefore given by
\begin{eqnarray}
\phi_--\phi_+=2(\theta_2-\theta_1),
\label{eq16}
\end{eqnarray}
having no parameter dependence except the relative polarization angle, $\theta_2-\theta_1$, between the two pulses. This is the difference between the Berry phases generated during the two time evolutions of $\ket{S_{1/2},\pm1/2}$, and, thus, the qubit rotation implemented by this geometric phase is robust against laser parameters as long as the adiabatic condition in Eq.~\eqref{adiacondition} is satisfied.
 
Now, we consider more general cases. First, when the dynamic Stark shift due to the off-resonant excited level, $P_{3/2}$, is taken into account, the detunings, $\Delta^{\pm}_{j}$ for $j=1,2$, are to be replaced by
\begin{widetext}
\begin{align}
&\Delta^+_j(t)=\Delta_j(t)+\frac{\abs{\bra{P_{3/2},1/2}\vec\mu\cdot\hat R\ket{S_{1/2},-1/2}E^+_j(t)}^2}{4\hbar^2\left(\omega_0+\Delta_{\rm fs}-\omega_p-2\Gamma_p \left(t-(-1)^j\frac{\tau}{2}\right)\right)}+\frac{\abs{\bra{P_{3/2},-3/2}\vec\mu\cdot\hat L\ket{S_{1/2},-1/2}E^-_j(t)}^2}{4\hbar^2\left(\omega_0+\Delta_{\rm fs}-\omega_p-2\Gamma_p \left(t-(-1)^j\frac{\tau}{2}\right)\right)} \\
&\Delta^-_j(t)=\Delta_j(t)+\frac{\abs{\bra{P_{3/2},3/2}\vec\mu\cdot\hat R\ket{S_{1/2},1/2}E^+_j(t)}^2}{4\hbar^2\left(\omega_0+\Delta_{\rm fs}-\omega_p-2\Gamma_p \left(t-(-1)^j\frac{\tau}{2}\right)\right)}+\frac{\abs{\bra{P_{3/2},-1/2}\vec\mu\cdot\hat L\ket{S_{1/2},1/2}E^-_j(t)}^2}{4\hbar^2\left(\omega_0+\Delta_{\rm fs}-\omega_p-2\Gamma_p \left(t-(-1)^j\frac{\tau}{2}\right)\right)},
\end{align}
\end{widetext}
where $\Delta_{\rm fs}$ is the fine-structure splitting of the excited states. However, since the linear-polarization condition guarantees $\Delta^+_j(t)=\Delta^-_j(t)$ for both $j=1,2$, so the presence of the $P_{3/2}$ makes no difference in Eq.~\eqref{eq16}. Second, when the polarization is not perfect, of non-zero ellipticity $\varepsilon_j$ ($j=1,2$), the condition $\abs{E^+_j(t)}=\abs{E^-_j(t)}$ is replaced by ${\abs{E^+_j(t)}^2}/{\abs{E^-_j(t)}^2}=({1+\varepsilon_j})^2/({1-\varepsilon_j})^2$. In this case,  the dynamic phases represented by the integrals in Eqs.~\eqref{eq13} and \eqref{eq14} in $\phi_+$ and $\phi_-$ are not equal, so the dynamic phase is to be included in the qubit rotation angle, making the gate sensitive to laser-parameter fluctuations.

\subsection{Experiment Data Analysis}

Figure 3 in the main text shows the experimental result of the $X$-rotation driven by two chirped pulses.  In Fig.~3(a), the measured $F=1$ state probability $\left|\left<1|\psi_{\rm final}\right>\right|^2$ is numerically fitted to the function 
\begin{equation}
P(\theta)=\gamma\sin^2(\theta+\Delta\theta)+\delta
\end{equation}
with fitting parameters $\gamma$, $\Delta\theta$, and $\delta$. The ideal case is $\gamma=1$, $\Delta\theta=\delta=0$, while
experimental imperfection results in degraded fringe visibility ($\gamma<1$ and $\delta>0$) and fringe shift ($\Delta\theta\neq 0$). Major errors are due to errors in state preparation and measurement (SPAM errors). In our experiment, there exist optical pumping infidelity ($\sim$4$\%$), push-out measurement infidelity ($\sim$3$\%$), and the polarization mismatch between the pulses and the quantization axis ($\sim$1$\%$), in addition to the effect of weak pre- and post-pulses~\cite{satellitepulse}. However, we note that the SPAM errors are not directly related to the robustness of the Berry-phase gate. 
On the other hand, nonzero fringe shift $\Delta\theta\ne0$ could imply failure of the intensity robustness of the proposed Berry-phase gate, but, as described in Sec.~II, it is attributed to the imperfect polarization due to the small birefringence in our optical setup~\cite{vacwindow}.

Figure~4 in the main text presents the result of the Ramsey interferometry using two pairs of chirped pulses to operate $ U_{\hat x}$ (pulse1 and pulse2) and $ U_{\hat n(T)}$ (pulse3 and pulse4), where $T$ is the time delay between the two pairs. The  $F=1$ state probability $| \bra1 U_{\hat n(T)}(2\theta)~U_{\hat x}(2\theta)\ket0 |^2$, measured as a function of the time delay variation $\Delta T$ from the initial time delay, is numerically fitted to the function
\begin{equation}
P(\theta, \Delta T)=\gamma_R \sin(2\theta) \cos^2(\pi f_R \Delta T + \phi_R)+\delta_R
\end{equation}
with fitting parameters Ramsey frequency $f_R$, Ramsey phase $\phi_R$, and fringe visibility and offset $\gamma_R$  and $\delta_R$, respectively. The measured frequency $\langle f_R \rangle =6.79\pm0.08$~GHz agrees well with the $^{87}$Rb hyperfine frequency $f_{\rm hf}=6.834682610904290(90)~{\rm GHz}$ within the 95$\%$ confidence interval, in which the equivalent time-domain error is as small as $0.9\pm1.8$~ps.

\subsection{Fidelity and robustness of ultrafast Berry-phase gates}

\begin{figure}[tbh]
\centerline{\includegraphics[width=0.48\textwidth]{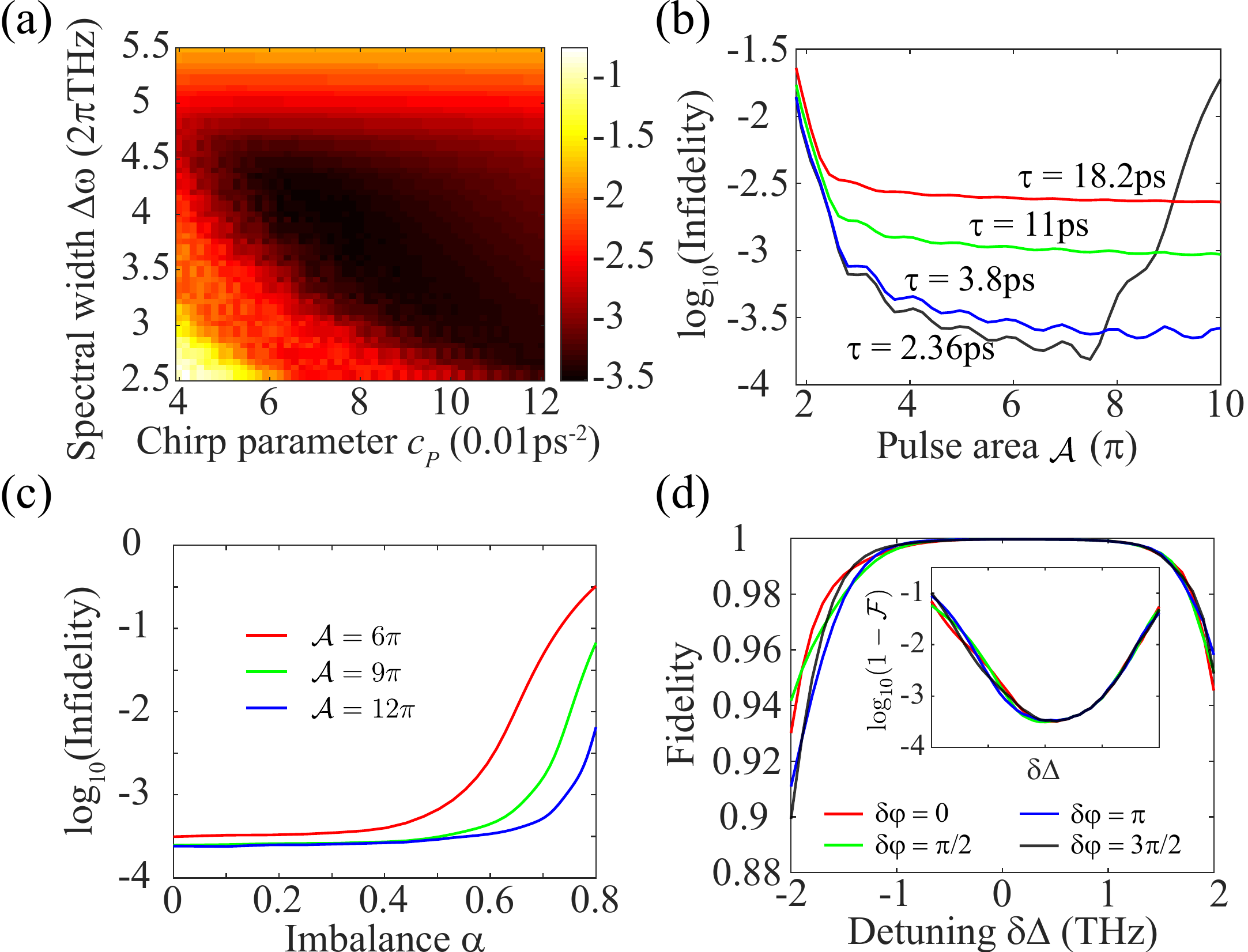}}
\hspace{\textwidth}
\begin{flushleft}
{FIG.~S1.  Calculated fidelity $\mathcal{F}$, for various laser parameters (either varied or otherwise fixed at $\Delta\omega=2\pi\times4~{\rm THz}$, $c_p=0.072~{\rm ps^{-2}}$, $\mathcal{A}=6\pi,\tau=4\Delta t_p$, $\alpha=0$, $\delta\Delta=0$, and $\Delta\phi=0$): (a) The infidelity $1-\mathcal{F}$ as a function of the spectral width $\Delta\omega$ and the chirp parameter $c_p$; (b) The infidelity $1-\mathcal{F}$ vs. the pulse area $\mathcal{A}$, for time delays $\tau=18.2$~ps (red), 11~ps (green), 3.8~ps (blue), and 2.36~ps (black); (c) The infidelity $1-\mathcal{F}$ vs. the amplitude imbalance $\alpha$, for pulse areas $\mathcal{A}=6\pi$ (red), $9\pi$ (green), and $12\pi$ (blue); (d) The fidelity $\mathcal{F}$ vs. the detuning $\delta\Delta$, for relative phases $\delta\varphi=0$ (red), $\pi/2$ (green), $\pi$ (blue), and $3\pi/2$ (black). (Inset  shows the log-scale infidelity.) }
\end{flushleft}
\label{suppfig1}
\end{figure}

We consider numerical estimation of  the fidelity and robustness of the given ultrafast Berry-phase gates. Lindblad master equation is used to calculate the amplitude and phase of the transition between the ground hyper-fine states $\ket{5S_{1/2},F=2,m_F}$ and $\ket{5S_{1/2},F=1,m_F}$, via $\ket{5P_{1/2},m_J=\pm1/2}\ket{I=3/2,m_I=\mp1/2}$ and $\ket{5P_{1/2},m_J=\pm1/2}\ket{I=3/2,m_I=\pm1/2}$, in the presence of the off-resonant coupling to $\ket{5P_{3/2}}$ and spontaneous decay.  The gate fidelity~\cite{gatefidelity} of ${U}_{\rm gate}$ is defined as 
\begin{equation}
\mathcal{F}={\abs{\bra{\psi_{\rm in}} {U}_{\rm ideal}^\dagger {U}_{\rm gate} \ket{\psi_{\rm in}}}^2},
\end{equation}
where ${U}_{\rm ideal}{={U}_{\hat x}(\pi)}$, ${U}_{\rm gate}={U}(\theta_2-\theta_1=\pi/2)$, and the result is averaged over the set of input states, i.e., $\ket{\psi_{\rm in}} \in \{\ket{0},\ket{1},(\ket{0}+\ket{1})/\sqrt{2},(\ket{0}+i\ket{1})/\sqrt{2}\}$. The contributing experimental parameters are the spectral width (FWHM) of the pulses, $\Delta\omega$, the chirp parameter, $c_p$, the pulse area, $\mathcal{A}=\int_{-\infty}^{\infty}dt\left(\Omega^+_1(t)+\Omega^+_2(t)\right)/2=\int_{-\infty}^{\infty}dt\left(\Omega^-_1(t)+\Omega^-_2(t)\right)/2$, the time delay, $\tau$, the amplitude imbalance of the two pulses, $\alpha=(\mathcal{E}_2 - \mathcal{E}_1)/(\mathcal{E}_2+\mathcal{E}_1)$, the frequency detuning, $\delta\Delta=\omega_1-\omega_0=\omega_2-\omega_0$, and the relative phase between the pulses, $\delta\varphi=\varphi_2-\varphi_1$, i.e., $\mathcal{F}=f(\Delta\omega,c_p,\mathcal{A},\tau,\alpha,\delta\Delta,\delta\varphi)$.

Figure~S1 shows the result of the calculation. The infidelity $1-\mathcal{F}(c_p,\Delta\omega)$ is shown in Fig.~S1(a). The high fidelity (low infidelity) region appears in the middle, upper-bounded by the leakage $D_2$ transition to $5P_{3/2}$ and lower-bounded by insufficient spectral width (smaller than required by the chirp), which corresponds to the spectral width ranged from about $2\pi\times3$~THz to $2\pi\times4$~THz in our experiment. As an example, we choose $\Delta\omega=2\pi\times4$~THz and $c_p=0.072$~ps$^{-2}$ for the rest of the calculation.

The robustness of the Berry-phase gates, against the laser power fluctuation, the pulse imbalance, and the frequency detuning, are respectively shown in Figs.~S1(b), S1(c), and S1(d).
First, in Fig.~S1(b), the infidelity $1-\mathcal{F}(\mathcal{A})$ is calculated for various time delays. The result exhibits fidelity plateaus, along which the Berry-phase gates are robust against the pulse area (or the laser power fluctuation). This $\mathcal{A}$-robust region is lower-bounded by nonadiabaticity and upper-bounded by the interference between temporally close two pulses. Next, the robustness against the amplitude imbalance ($\alpha$-robustness) is shown in Fig.~S1(c), where a sufficiently large pulse area ensures the $\alpha$-robustness because the adiabatic condition of the chirped RAP breaks down for a weaker pulse of insufficient Rabi frequency. 
Finally, the robustness against the detuning $\delta\Delta$ and relative phase $\delta\varphi$ is shown in Fig.~S1(d). The $\delta\Delta$-robustness is achieved around the zero detuning,  regardless of any relative phase ($\delta\varphi$-robustness), while the asymmetry between positive and negative detunings stems from the dynamic Stark shift (due to the D$_2$ transition) of $5S_{1/2}$ level. The fidelity above 0.999 can be achieved in the range as wide as the laser spectral width in our experiment.

\end{document}